\documentclass[aps,prb,twocolumn,showpacs,floats,floatfix]{revtex4}

\usepackage{epsfig}
\usepackage{amsmath}
\newcommand{\be}{\begin{equation}}
\newcommand{\ee}{\end{equation}}
\newcommand{\bea}{\begin{eqnarray}}
\newcommand{\eea}{\end{eqnarray}}
\newcommand{\ba}{\begin{array}}
\newcommand{\ea}{\end{array}}
\newcommand{\nn}{\nonumber}
\newcommand{\parti}{\partial}

\newcommand{\del}{\delta}

\newcommand{\al}{\alpha}
\newcommand{\sig}{\sigma}

\newcommand{\eps}{\epsilon}
\newcommand{\lam}{\lambda}
\begin{document}

\title{\bf Quantum Pumping and Quantized Magnetoresistance in a Hall Bar} 

\author{M. Blaauboer
}

\affiliation{ Department of NanoScience, Delft University of Technology,
Lorentzweg 1, 2628 CJ Delft, The Netherlands}
\date{\today}

\begin{abstract}
We show how a dc current can be generated in a Hall bar without applying a bias voltage. 
The Hall resistance $R_H$ that corresponds to this pumped current is quantized, just 
as in the usual integer quantum Hall effect (IQHE). In contrast with the IQHE, 
however, the longitudinal resistance 
$R_{xx}$ does not vanish on the plateaus, but equals the Hall resistance. We propose 
an experimental 
geometry to measure the pumped current and verify the predicted behavior of $R_H$ and $R_{xx}$. 
\end{abstract}

\maketitle

\section{Introduction}

The phenomenon of quantum pumping has received a lot of attention in recent years\cite{spiv95,brou98,swit99}.
 It involves the generation of a dc current of electrons through ac modulation of their environment, 
e.g. the potential landscape or a magnetic field. This requires that at least two independent system 
parameters are varied in a periodic manner, and with a phase difference $\phi$ between them\cite{thou83}. 
If the resulting current depends on quantum-mechanical interference in the system, this process is 
named ``quantum pumping'', in contrast with ``classical'' pumping in which particles are pumped 
sequentially\cite{kouw94}. So far, quantum pumping has mainly been studied in the context of 
charge\cite{levi01} and spin\cite{mucc02} pumping in quantum dots, small islands embedded in 
a semiconductor material\cite{kouw97}. In this work, we focus on a new pumping geometry: a Hall 
bar in the integer quantum Hall regime, see Fig.~\ref{fig:Hallbar}. We recently proposed how 
quantum pumping in such a system may be used to control and measure the local polarization of the 
nuclei present in the system\cite{blaa03}. Here we study in more detail the pumped current 
itself and investigate the corresponding perpendicular (Hall) and parallel resistance. 
Using a Floquet scattering approach, we show that in the regime of weak pumping 
and low temperatures the pumped current exhibits typical pumping characteristics such 
as linear dependence on the pumping frequency $\omega$, quadratic dependence on the pumping 
strength and sinusoidal dependence on the phase 
difference $\phi$. For stronger pumping, numerical analysis shows that the latter disappears, 
although the periodicity in $\phi$ remains, and that the current becomes more sharply peaked. 
We then calculate the Hall resistance and find that it exhibits steplike behavior as a 
function of the applied magnetic field. We also calculate the longitudinal resistance 
$R_{xx}$ which turns out to be identical to $R_H$. This can be explained in terms of 
the intrinsic scattering nature of quantum pumping.

\section{Quantum pumping}

\begin{figure}
\centerline{\epsfig{figure=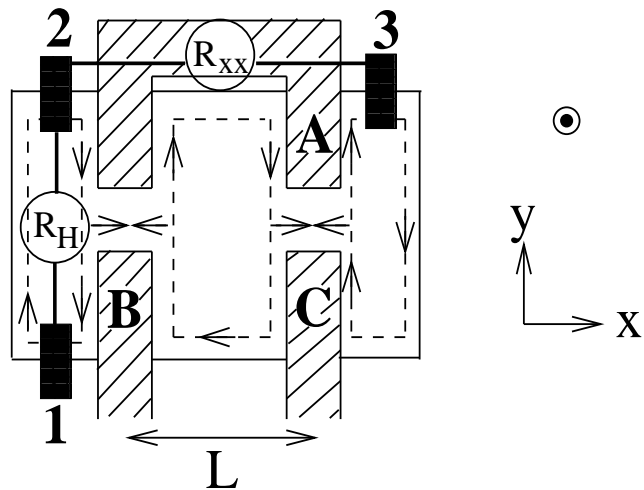,height=6.4cm,width=0.98\hsize}}
\caption[]{Schematic picture of a Hall bar at bulk filling factor $\nu=1$ with a 
magnetic field applied perpendicular to the plane of the paper. Current (dashed lines) 
flows along the edges of the sample and is scattered at the two quantum point 
contacts AB and AC (shaded). See the text for further explanation.
}
\label{fig:Hallbar}
\end{figure}

Let us start by considering the geometry schematically depicted in Fig.~\ref{fig:Hallbar}. 
%
%
It consists of a Hall bar placed in a sufficiently strong magnetic field, so that electrons 
travel in one-dimensional channels along the boundaries of the sample\cite{been91}. 
These so-called edge channels correspond to quantized energy levels, the Landau levels, which are 
macroscopically degenerate, capable of holding many electrons. The number of filled 
Landau levels is characterized by the filling factor $\nu\equiv n_e h/(eB)$, where 
$n_e$ denotes the electron density and $B$ the applied magnetic field. The Hall bar 
in Fig.~\ref{fig:Hallbar} contains three contacts 1-3 which can be used to measure 
the voltage drops perpendicular and parallel to the direction of the current. 
It also contains three voltage gates A, B, and 
C which, combined as the split-gate pairs AB and AC, form two quantum point contacts (QPCs). 
By applying time-varying voltages with a phase difference $\phi$ to gates B and C, 
these gates may serve as pumping parameters. To begin with, we 
concentrate on the regime of $\nu=1$ and assume the QPCs transmit at most one edge 
channel. Part of the electrons in the edge channel are then reflected at, and part 
of them transmitted through each QPC, as depicted in Fig.~\ref{fig:Hallbar}. Since transport 
is effectively one-dimensional, we model the QPCs by $\del$-function potential 
barriers\cite{enti01}  $V_{\rm AB}(x,t)= \left[ \bar{V}_{\rm AB} + 2\, \del V_{AB} 
\cos (\omega t) 
\right]\, \del (x+\frac{L}{2})$ and $V_{\rm AC}(x,t)= \left[ \bar{V}_{\rm AC} + 2\, 
\del V_{AC} \cos (\omega t + \phi) \right]\, \del (x-\frac{L}{2})$.
The time-dependent parts represent the pumping voltages, applied to gates B and C, 
with $\omega$ the pumping frequency and $\phi$ a phase difference\cite{2}.
In order to calculate the pumped current, we use a Floquet scattering approach
\cite{shir65,mosk02}. This approach describes the scattering of electrons at an 
oscillating barrier in terms of the gain or loss of energy quanta $\hbar \omega$. 
If an incoming electron has energy $E$, the outgoing state after interaction 
with the barrier is characterized by energies $E_n \equiv E + n\hbar \omega$, 
with $n=..,-2,-1,0,1,2,..$. The Floquet theorem states that this is the full 
set of energies for outgoing particles\cite{shir65} and the scattering process
 is represented by the Floquet scattering matrix $S_{\rm FL}$. The matrix elements 
$t_{\al \beta}(E_n,E)$ of $S_{\rm FL}$ represent the scattering amplitude of 
an electron that arrives at the barrier from contact $\beta$ with energy $E$ 
and leaves to contact $\al$ with energy $E_n$. The pumped current into contact 
$\al$, with in our case $\al =1,2$ or 3, is then given by\cite{mosk02}
\bea
I_{\al} & = & \frac{e}{h} \int dE f(E) \sum_{\beta} \sum_{E_n >0}
\left( |t_{\al \beta}(E_n,E)|^2 - \right. \nn \\
& & \left. \hspace{2.3cm} |t_{\beta \al}(E_n,E)|^2 \right).
\label{eq:current}
\eea
Here $f(E)=(1+ {\rm exp}[(E-\mu)/(k_B T)])^{-1}$ denotes the Fermi distribution, 
with $\mu$ the chemical potential in the leads, and we have assumed that incoming 
electrons in different leads are described by the same Fermi function. This is in 
agreement with our pumping set-up, where no bias voltage is applied and 
all contacts are held at the same chemical potential $\mu$. Under these circumstances, current flows out of and into each contact 1, 2 and 3, as sketched in Fig.~\ref{fig:Hallbar}: all electrons emerging from contact 1 will flow into contact 2, the ones emerging from contact 2
are scattered at the QPC's and go to contact 1 or 3, and electrons coming out of contact 3 
are either scattered back into 3 or end up in contact 1. We are interested in the net current flow $I_{\al}$ into each contact {$\al$ = 1,2,3}. It is easy to see that $I_2=0$, since the amount of 
incoming and outgoing electrons at contact 2 is the same.
So, in order to calculate the pumped current, it is sufficient to consider a 
two-terminal geometry, omitting contact 2, and
\bea
I_1 = - I_3 & = & \frac{e}{h} \int dE f(E) \sum_{E_n >0}
\left( |t_{13}(E_n,E)|^2 - \right. \nn \\
& & \hspace{2.3cm} \left. |t_{31}(E_n,E)|^2 \right).
\label{eq:current1}
\eea
Now it remains to calculate the scattering amplitudes. This is done by matching 
the appropriate wavefunctions at the two $\del$-function barriers, as was recently
 done for one-dimensional mesoscopic quantum pumps\cite{mosk02,zhu02}. The 
wavefunction describing electron transport along the edge channels is given 
by the solution of the Schr\"odinger equation $i\hbar \frac{\parti}{\parti t} 
\psi(x,y,t) = {\cal H} \psi(x,y,t)$ with
\be
{\cal H} = \frac{1}{2m^{*}} (i\hbar \vec{\nabla} + e \vec{A})^2 + V(y).
\label{eq:Ham}
\ee
Here $m^{*}$ denotes the effective mass, $\vec{A}$ the vector potential and $V(y)$ 
the transverse confining potential\cite{spineffects}. We now take the Landau gauge 
$A_x=-By$ and assume that the transverse confining potential can be described by 
a harmonic potential, $V(y) = \frac{1}{2} m^{*} \omega_0^2 y^2$. The latter gives a good 
description of transport at the edges if $V(y)$ is constant on the scale of the 
magnetic length $l_m \equiv 
\left( \frac{\hbar}{m^{*}\sqrt{\omega_0^2 + \omega_c^2}}\right)^{1/2}$, with 
$\omega_c\equiv |e|B/m^{*}$ the cyclotron frequency\cite{jans94,hale99}. The solution of the
Schr\"odinger equation with the Hamiltonian (\ref{eq:Ham}) is then given by\cite{laug98}
\bea
\psi_{\lam}(x,y,t) &=& C_{\lam}\, \chi_{\lam}(y)\, e^{i(kx - \frac{E_{\lam}}{\hbar}t)}, 
\label{eq:wavefunction} \\
\chi_{\lam}(y) & = & e^{\frac{1}{2 l_m^2}(y+\frac{\omega_c^2}{\omega_0^2+\omega_c^2} 
y_k)^2} \times \nn \\
& & \ \ \ \left( \frac{\parti}{\parti y} \right)^{\lam} e^{-\frac{1}{l_m^2}(y+
\frac{\omega_c^2}{\omega_0^2+\omega_c^2} y_k)^2} \\
E_{\lam} & = & (\lam + \frac{1}{2}) \hbar \sqrt{\omega_0^2 + \omega_c^2} + 
\frac{1}{2} m^{*} \frac{\omega_{0}^2 \omega_c^2}{\omega_{0}^2 + \omega_c^2}
y_{k}^2, \nn \\
& &\ \ \ \ {\rm for}\ \   \lam = 0,1,2,... 
\label{eq:LLs}
\eea
Here $C_{\lam}$ denotes a normalization constant, $y_{k} \equiv \hbar k/eB$, and 
$E_{\lam}$ the Landau levels 
measured from the lowest band imposed by the confinement in the $z$-direction. When the
two time-dependent $\del$-function potentials are included into the Hamiltonian (\ref{eq:Ham}), the
Floquet theorem says that the eigenstates of this new Hamiltonian can be represented as a
superposition of the wavefunctions (\ref{eq:wavefunction}) with shifted energies:
\bea
\psi_{FL}(x,y,t) = \sum_{n=-\infty}^{\infty} \psi_{\lam,n}(x,y)\, e^{-i n \omega t} e^{-i 
\frac{E_{\lam}}{\hbar} t},
\label{eq:wavefunction1}
\eea
with 
\bea
\psi_{\lam,n}(x,y) & = & C_{\lam}\, \chi_{\lam}(y)\, e^{i k_n x},
\eea
\bea
k_n & = & \sqrt{\frac{2m^{*}}{\hbar^2} \frac{\omega_0^2+\omega_c^2}{\omega_0^2}} 
\left[ E_{F} + n\hbar \omega - \frac{1}{2} \hbar \sqrt{\omega_0^2 + \omega_c^2} 
\right]^{1/2}.
\label{eq:wavevector}
\eea
The wavevector (\ref{eq:wavevector}) applies for the lowest Landau level. The 
scattering amplitudes are obtained by matching (\ref{eq:wavefunction1}) at the 
boundaries $x=-\frac{L}{2}$ and $x=\frac{L}{2}$. For an electron coming from the 
left the wavefunctions to be matched are given by
\bea
\psi_{\rm FL}(x < -\frac{L}{2}) & = & e^{ik_0x} + \sum_{n=-\infty}^{\infty} 
r_n e^{-ik_nx} e^{-in\omega t} \\
\psi_{\rm FL}(-\frac{L}{2} < x < \frac{L}{2}) & = & \sum_{n=-\infty}^{\infty} 
\left( A_n e^{i k_n x} + B_n e^{-i k_n x} \right) \times \nn \\
& & \hspace{2cm}  e^{-i n \omega t} \\
\psi_{\rm FL}(x > \frac{L}{2}) & = & \sum_{n=-\infty}^{\infty} t_n e^{i k_n x}
e^{-i n \omega t}.
\eea
Similar wavefunctions apply for electrons incident from the right. Here we have omitted
the time-dependent factor $e^{-i \frac{E_0}{\hbar} t}$ and transverse wavefunction 
$\chi_1(y)$, since they are common to all wavefunctions. Following the same procedure 
as in Appendix A of Ref.~\cite{mosk02}, we arrive at the result
\begin{subequations}
\bea
r_n & = & (A_n - \del_{n,0}) e^{-i k_n L} + B_n \\
t_n & = & A_n + B_n e^{-i k_n L},
\eea
\end{subequations}
where $A_n$ and $B_n$ are given by recursion relations (Eqs.~(A10)-(A16) in 
Ref.~\cite{mosk02} with $k_n$ given by Eq.~(\ref{eq:wavevector})). We have solved 
these equations and calculated the resulting current numerically. Before discussing 
this result, let us look at the special case of weak pumping, for which an analytic 
solution can be obtained. Weak pumping is characterized by $\del V_{\rm AB(AC)} 
\ll \bar{V}_{\rm AB (AC)}$. Assuming, for simplicity, equal QPCs with $\bar{V}_{\rm AB} 
= \bar{V}_{\rm AC} \equiv \frac{\hbar^2}{m^{*}} p$ and
$\del V_{\rm AB} = \del V_{\rm AC} \equiv \frac{\hbar^2}{m^{*}} q$, this translates 
into the requirement $q \ll k_0$. In this case it is easy to show that the scattering probability 
into the sideband with energy $E_n = E+n\hbar \omega$ is proportional to $\left( \frac{q^2}{k_0^2} 
\right)^{|n|}$, so that in good approximation only the lowest sideband for $n=\pm 1$ 
has to be taken into account. We then set $A_{\pm 2} = B_{\pm 2}=0$ in the recursion 
relations of Ref.~\cite{mosk02} and calculate $A_0$, $A_{\pm 1}$, $B_0$ and $B_{\pm 1}$. 
For nearly-open QPCs, with $p \ll k_0$, this yields the transmission amplitudes, to 
lowest order in $q/k_0$,
\begin{subequations}
\bea
t_{31,0} & = & 1 + 2i \frac{q^2}{k_0^2} \sin (k_0 L) \left( e^{i\phi} e^{i k_1 L} + 
e^{-i \phi} e^{i k_{-1}L} \right) \\
t_{31,\pm 1} & = & -i \frac{q}{k_0} \left( e^{-i(k_0-k_{\pm 1}) \frac{L}{2}} + 
e^{\mp i \phi}  e^{i(k_0-k_{\pm 1}) \frac{L}{2}} \right) \\
t_{13,0} & = & 1 + 2i \frac{q^2}{k_0^2} \sin (k_0 L) \left( e^{-i\phi} e^{i k_1 L} + 
e^{i \phi} e^{i k_{-1}L} \right) \\
t_{13,\pm 1} & = & -i \frac{q}{k_0} \left( e^{-i(k_0-k_{\pm 1}) \frac{L}{2}} + 
e^{\pm i \phi}  e^{i(k_0-k_{\pm 1}) \frac{L}{2}} \right). 
\eea
\end{subequations}
The resulting current is given by
\begin{subequations}
\bea
I_3 &  =  & - I_1 \nn \\
& = & - 8 \frac{e}{h} q^2 \sin \phi \int dE f(E) \frac{1}{\tilde{k}_0^2} 
\cos (2 \tilde{k}_0 L)
\sin (\eps \tilde{k}_0 L) \\
& \approx & -8 \frac{e}{h} \mu \frac{q^2}{k_0^2} 
\sin \phi \cos (2 k_0 L)
\sin (\eps k_0 L) \hspace{0.15cm} {\rm for}\ k_B T \ll \mu,
\label{eq:curr2appr}
\eea
\label{eq:current2}
\end{subequations}
with 
\bea
\epsilon & \equiv & \frac{1}{2}\, \frac{\hbar \omega}{E_F - \frac{1}{2} \hbar 
\sqrt{\omega_0^2 + \omega_c^2}} \hspace{1cm} {\rm and} \nn \\
\tilde{k}_{0} & = & k_{0} (E_F \rightarrow E) \nn
\eea
For $\epsilon k_0 L \ll 1$ the current (\ref{eq:curr2appr}) is  directly proportional 
to the pumping frequency $\omega$. In the derivation of Eq.~(\ref{eq:current2}) we 
have assumed that $\epsilon \ll 1$, which is true for typical system parameters, 
see Sec.~\ref{sec:discussion}. In this case $k_n = k_0 \sqrt{1 + 2n\eps} \approx 
k_0 + k_0 n \eps$ for $n$ not too large ($n\eps \ll 1$), so certainly in the 
situation of weak pumping. One may then use this expression of $k_n$ in the 
exponents and set $k_n \approx k_0$ in the prefactors. The latter is analogous 
to the Andreev approximation which is often made in calculations involving NS 
interfaces\cite{andr64}.

\begin{figure}
\centerline{\epsfig{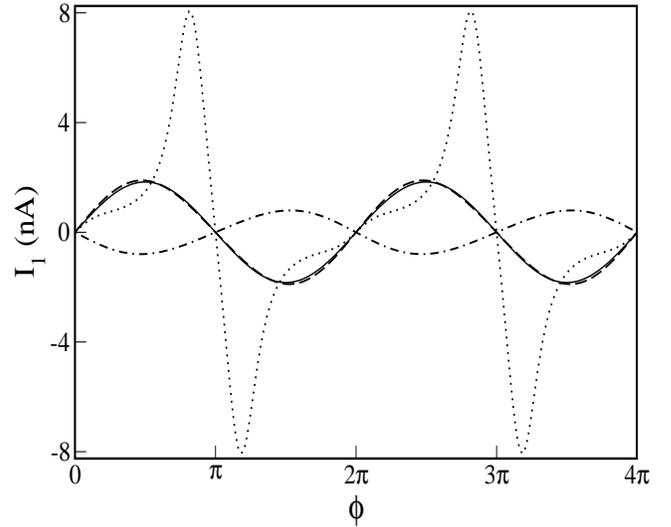}}
\caption[]{The current $I_1$ pumped into contact 1 as a function of the 
phase difference $\phi$. Parameters used are : $B=7\, T$, $m^{*}= 0.61\ 10^{-31} 
kg$ (for GaAs), $\omega_c = 1.836\ 10^{13} s^{-1}$, $\omega_0 = 0.08\, \omega_c$\cite{hale99}, 
$E_F=1.716\ 10^{-21}$ J (for $n_e = 3\, \, 10^{15} m^{-2}$), $L=300\, nm$\cite{elze02}, 
and $\omega = 10^8 s^{-1}$. The solid line represents the analytical result 
Eq.~(\ref{eq:curr2appr}), and the other lines represent numerical results for
 $p/k_0=1/10$, $q/k_0=1/6$ (dashed line), $p/k_0=7/6$, $q/k_0=1/6$ (dot-dashed line), 
$p/k_0=1/10$, $q/k_0=1$ (dotted line).
}
\label{fig:current}
\end{figure}

Fig.~\ref{fig:current} shows the pumped current $I_1$ as a function
of the phase difference $\phi$ at low temperatures in the weak and strong pumping
limits.
%
%
As expected, in the weak pumping limit and for nearly-open QPCs the numerical 
result (dashed line) agrees very well with the analytical one (solid line). 
The current is proportional to $\sin \phi$ and we have checked that it displays 
other typical pumping characteristics such as being linearly proportional to 
both the frequency $\omega$ and the pumping amplitude squared ($q^2$)\cite{direction}. Also 
for less open QPCs (dot-dashed line for which the transmission probability of 
both QPCs $T_{\rm AB}=T_{\rm AC} \approx 0.57$) the pumped current behaves 
in the same way, the main difference being that it now has a smaller 
amplitude. For higher pumping strength the current acquires a more complex 
structure (dotted line). It is not proportional to $\sin \phi$ anymore, 
although it is still periodic. For the set of parameters used here, the current 
spreads over $\pm$ 40 sidebands upon scattering at each QPC, many more than 
in the weak pumping regime.

\section{Hall and Longitudinal Resistance}

Having found that a current can be pumped across a Hall bar in the absence 
of a bias voltage, it is interesting to ask what resistance corresponds to 
this current. In the usual integer quantum Hall effect a current is 
driven through a Hall bar at low temperatures and gives rise to a resistance 
parallel to the direction of the current (the longitudinal resistance $R_{xx}$) 
and a resistance perpendicular to the direction of the current (the Hall 
resistance $R_H$)\cite{Hallres}. When measured as a function of the applied 
magnetic field, the Hall resistance exhibits plateaus at integer multiples 
of $h/e^2$, a new plateau appearing when the number of filled Landau levels
 changes by one. The explanation for the appearance of the plateaus, on 
which the resistance does not change, lies in the presence of disorder 
in the system. This leads to the formation of localized states which in 
turn lead to the quantization because, roughly speaking, when the Fermi 
energy moves through an energy band of localized states the number of 
conduction electrons and hence the Hall resistance does not change. For 
the same reason, the longitudinal resistance vanishes on the plateaus 
since the localized electrons cannot dissipate energy by making a transition
 to a state that is lower in energy. When a new Landau level is occupied 
(or emptied) and the Hall resistance jumps to the next plateau, the 
longitudinal resistance exhibits a peak before returning to zero.

How, then, is the behavior of $R_H$ and $R_{xx}$ if the current is not driven 
through the sample but pumped, as described in the above? Is the Hall 
resistance then also quantized and does the longitudinal resistance also 
vanish? Studying the configuration of Fig.~\ref{fig:Hallbar} we will argue 
in the following that the answer to the first question is yes and to the 
second question is no.

Consider again the Hall bar of Fig.~\ref{fig:Hallbar} in the regime of $\nu=1$ 
and assume that the QPC's transmit at most one edge channel during the 
entire pumping cycle, so the conductance $G_{\rm AB}$, $G_{\rm AC}$ $\leq \frac{e^2}{h}$. 
In the absence of scattering at the QPC's
all three terminals are at the same chemical potential $\mu$. In order to 
calculate $R_{xx}$ and $R_H$ we first look at the voltage drop across 
terminals 1 and 2 (for $R_H$) and 2 and 3 (for $R_{xx}$). Let $R_{\al \beta}= 
\sum_{n} |r_{\al \beta}|^2$ and $T_{\al \beta} = \sum_{n} |t_{\al \beta}|^2$ 
denote the reflection and transmission 
probabilities from contact $\beta$ to contact $\al$. Analyzing incoming 
and outgoing currents at zero temperature, the three contacts are then 
characterized by the chemical potentials\cite{datt95}
\bea
\mu_1 & = & (R_{12} + T_{13} ) \mu \nn \\
\mu_2 & = & \mu \nn \\
\mu_3 & = & (R_{33} + T_{32} ) \mu. \nn 
\eea
These lead to the voltage drops
\bea
V_H \equiv V_2 - V_1 & = & (1 - R_{12} - T_{13}) \frac{\mu}{e} = 
(T_{32} - T_{13}) \frac{\mu}{e} \nn \\
V_{xx} \equiv V_3 - V_2 & = & (T_{32} + R_{33} - 1) \frac{\mu}{e} =
(T_{32} - T_{13}) \frac{\mu}{e}. \nn
\eea
Here we have used the conservation laws $T_{13} + R_{33} =1$ and $R_{12} + T_{32}=1$.
It thus turns out that the parallel and perpendicular voltage drops are the same, 
$V_{H}=$ $V_{xx} \equiv$ $V$, and hence that the Hall resistance and 
longitudinal resistance are equal:
\bea
R_H = R_{xx} = \frac{V}{I_3} = \frac{h}{e^2}.
\label{eq:Hallres}
\eea
In the last step of (\ref{eq:Hallres}) we have used Eq.(\ref{eq:current1}) for the 
current $I_3$ at low temperatures $k_B T \ll \mu$. The same result would also be 
obtained at higher temperatures. Eq.~(\ref{eq:Hallres}) is valid 
on a plateau, when the Fermi energy lies in a subband of 
localized states. The reason why $R_H$ is quantized is the same as in the usual 
IQHE, and due to the presence of impurities. More surprising, at first sight, 
is the quantization of $R_{xx}$ and it being equal to $R_{H}$. When looking 
more closely at Fig.~\ref{fig:Hallbar}, the reason for this becomes clear and 
lies in the intrinsic scattering nature of quantum pumping. The presence of 
scattering, here in the form of the two QPC's, is essential to obtain a pumped 
current: in the absence of the QPC's no such current would flow. Moreover, 
even in the presence of the QPC's but if they are wider and transmit more 
than 1 edge channel, no current would flow: the time-dependent voltages 
would periodically push the edge channel a bit more inwards, so that the
electrons flow along and around the barriers imposed by the voltage leads, and no 
net current would be pumped into contact 1 or 3. In short, to obtain a 
pumped current it is required that (part of) the electrons are scattered. 
As a result, dissipation also occurs in the direction parallel to the current flow and 
hence $R_{xx}$ never drops to zero. This was also found for the usual IQHE, 
if the Hall bar contains one or more QPCs\cite{jans94}. 
At the same time, due to the absence of a bias voltage, 
the difference between parallel and perpendicular resistance vanishes\cite{pumping}. Each outgoing 
electron from contact 2 or 3 is scattered into either contact 1 or 3. Since the 
outgoing electrons are all at the same chemical 
potential and the current is conserved, both $V_H$ and $V_{xx}$ only depend 
on the difference
between the transmission probabilities from 1 to 3 and vice versa. 

\begin{figure}
\centerline{\epsfig{figure=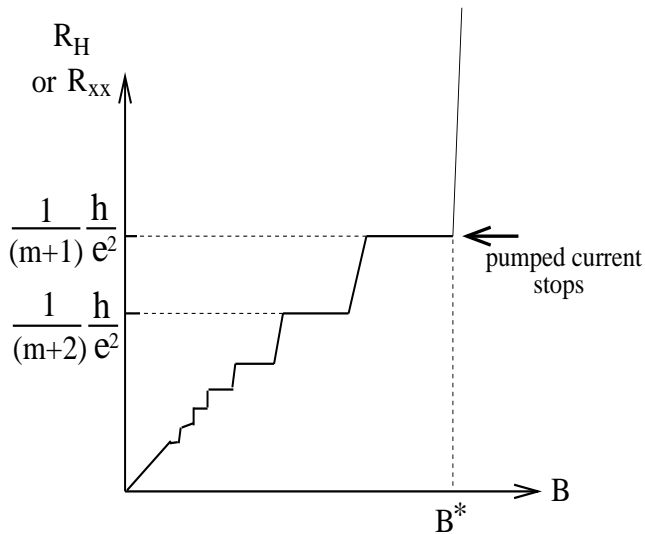,height=7.0cm,width=0.98\hsize}}
\caption[]{Schematic picture of the Hall resistance $R_H$ and longitudinal 
resistance $R_{xx}$ as a function of 
the applied magnetic field $B$. 
At $B^{*}$ the (m+1)th Landau level becomes occupied, 
with m the highest Landau level which is still fully transmitted by the QPC's.
}
\label{fig:resistances}
\end{figure}

So far, we have considered filling factor $\nu=1$ and QPC's with transmission less 
than unity. What happens at higher filling factors and different transmission of the 
QPC's? Let $\nu=n$ and assume the QPC's, still taken to be equal, transmit $m$ edge 
channels (with $n$ and $m$ both integers) plus a fraction $T$, with $0\leq T\leq 1$, 
of the (m+1)th channel. Then two situations can be distinguished:
\bea
1) & n\leq m & \rightarrow  {\rm no\ pumping\ takes\ place,\ all\ edge\ channels} \nn \\
&& \hspace{0.45cm} {\rm are\ transmitted\ without\ scattering.} \nn \\
2) & n>m & \rightarrow  {\rm The\ (m+1)th\ edge\ channel\ gives\ rise} \nn \\
&& \hspace{0.45cm} {\rm to\ a\ pumped\ current\ and\ corresponding} \nn \\
&& \hspace{0.45cm} {\rm resistances}\ R_H = R_{xx} = \frac{h}{ne^2}. \nn
\eea
As a function of decreasing magnetic field, or equivalently increasing filling factor 
$\nu$, the Hall resistance thus rises from 0 to the value $\frac{h}{(m+1)e^2}$ as 
soon as the (m+1)th Landau level becomes occupied at magnetic field $B^{*}$. This
is depicted in Fig.~\ref{fig:resistances}. 
For higher filling factors the resistance
 decreases in steps, 
just as in the usual IQHE.
%
%
Similar arguments apply if the transmission probabilities of the two QPC's are not equal. 

\section{Discussion and Conclusions}
\label{sec:discussion}

In this final section, let us look at realistic experimental parameters in order to 
quantify our results. Typical Hall bars are several 100 $\mu$m long and about 100 $\mu$m 
wide\cite{jans94}. Present-day experimental techniques allow for separation of the gates by 
200-300 nm\cite{elze02}, so that the QPC gates can be placed sufficiently far away from the 
other contacts 1-3 that the applied time-varying voltages do not affect the voltage 
measurements there. For the parameters used in Fig.~\ref{fig:current} (see the caption)
and temperatures $T\sim 10$ mK the conditions $k_B T \ll \mu$ and $\epsilon \ll 1$
are fulfilled and a pumped current on the order of 1-10 nA would be generated, well 
within reach of observation. 
Our model does not take dephasing effects into account. These are detrimental to 
quantum pumping, but were found to be small\cite{swit99} at temperatures 
below 100 mK.

In conclusion, we have found that a dc current can be pumped across a Hall bar  
containing two QPC's without applying a bias voltage. Just like a current that is driven 
through the sample, this pumped current gives rise to a Hall and longitudinal resistance. 
In sharp contrast with the usual IQHE, however, the longitudinal resistance equals the 
Hall resistance and both exhibit plateaus as a function of the applied magnetic field. 
A cut-off occurs when the filling factor falls below the number of Landau levels 
transmitted through the QPC's. Measurement of $R_H$ and/or $R_{xx}$ as a function of 
the magnetic field can thus be used to deduce $\nu$ and the transmission of the QPC's.
Also the sharp change in resistance at $B^{*}$ may be useful, e.g. as a magnetic switch.
We hope that these curious properties of quantum pumping in high magnetic fields will 
find experimental confirmation.

Stimulating discussions with M. Heiblum and Y. Levinson are gratefully acknowledged. 
This work has been 
supported by the Stichting voor Fundamenteel Onderzoek der Materie (FOM), and by 
the EU's Human Potential Research Network under contract No. 
HPRN-CT-2002-00309 (``QUACS'').


\begin{thebibliography}{99}
\bibitem{2}
The factor of 2 in front of the cosines is introduced for convenience.
\bibitem{spineffects}
If spin effects are important, additional terms have to be included
in the Hamiltonian (\ref{eq:Ham}), such as the Zeeman term 
$\frac{1}{2} g^{*} \mu_B B \sig$ [with $g^{*}$ the effective g-factor, 
$\mu_B$ the Bohr magneton and $\sig = 1\, (-1)$ for electrons with 
spin up (down)] and, for example,
a hyperfine or spin-orbit interaction term.
\bibitem{direction}
In contrast to quantum pumping in quantum dots, however, the direction of
the pumped current in high magnetic fields, such as quantum Hall systems,
is fully determined.
\bibitem{pumping}
We emphasize that it is thus the fact that current is pumped, and not the details of the 
geometry, that causes $R_H$ and $R_{xx}$ to be equal.
\end{thebibliography}
\end{document}